# Engineering graphene by oxidation:

# a first principles study

Zhiping Xu [1*] and Kun Xue [2]

[1]Department of Civil and Environmental Engineering, Massachusetts Institute of Technology, Cambridge 02139, MA, USA

[2]Department of Engineering Mechanics, Tsinghua University, Beijing 100084, China

[*]Correspond author, email: xuzp@mit.edu

*Abstract*:

Graphene epoxide, with oxygen atoms lining up on pristine graphene sheets, is investigated theoretically in this Letter. Two distinct phases: metastable *clamped* and *unzipped* structures are unveiled in consistence with experiments. In the stable (unzipped) phase, epoxy group breaks underneath $sp^2$ bond and modifies the mechanical and electronic properties of graphene remarkably. The foldable epoxy ring structure reduces its Young's modulus by 42.4%, while leaves the tensile strength almost unchanged. Epoxidation also perturbs the π state and opens semiconducting gap for both phases, with dependence on the density of epoxidation. In the unzipped structures, localized states revealed near the Fermi level resembles the edge states in graphene nanoribbons. The study reported here paves the way for oxidation-based functionalization of graphene-related materials.

*Introduction*

Recently developed techniques to isolate single graphene layer [1] provides the possibility to investigate this extremely mono-layer materials. With a hexagonal graphitic lattice, this material holds outstanding properties such as massless Dirac fermions[2], abnormal quantum Hall effects [3], and extremely high stiffness and structural stability[4]. Besides of the intrinsic properties of graphene, fresh physics has also been revealed in graphene-related nano-structures, such as finite width graphene nano-ribbons with half metalicity [5] and patternable transport properties [6, 7] have been reported. One of the most convenient processes to obtain graphene materials is chemical exfoliation [8-10], where graphite crystals were first exfoliated through oxidation, then reduced into graphene monolayers. This method has the advantage of massive production and possibility of deposition from solution [9, 11] in comparison with other approaches like mechanical cleavage [1], epitaxial growth [12] and organic synthesis [13]. However, the products of current technique contain considerable amount of the intermediate product – graphene oxides, with a large number of epoxy and hydroxyl groups [14]. The incomplete reduction leaves wrinkles and folds in the graphene sheet products and is detrimental to their electronic properties [15].

Interests have also arisen for functional graphene materials such as graphene oxides [16-20]. Although have been studied for decades, there are very limited knowledge on the atomic structure and the physical properties of the graphene oxides. Different atomic models have been proposed with various functional groups [17]. The most notable structure proposed is the Lerf-Klinowski model on the basis of NMR spectroscopy data [20], which contains epoxy and hydroxyl groups on the graphene sheets or at their edges. Fault lines in graphene have been observed experimentally as a result of oxidation [18]. It was explained as adsorbed epoxy groups that tend to lined up cooperatively. Recent ultra-high vacuum scanning probe microscopy study revealed locally periodic structures. The atomic structure of lattice was identified to be oxygen atoms bound to the graphene sheet [19]. These oxidized monolayer sheets are structurally close to graphene and are expected to have similarly properties. One can thus expect appealing functional graphene structures in a controllable manner through chemical oxidation and reduction [17, 21]. These features provide graphene epoxide materials great potential for applications. For this purpose, a good understanding on their structural and electronic properties are desired. In this work, we will discuss their energetic, structural and electronic properties, based on first principles calculations. Although experimental observation [22] concluded that graphite oxides are globally disordered non-stoichiometric phases containing both epoxide and hydroxyl groups, we hope the structural-properties relations presented here for regularly patterned structures can shed some light on possible applications of this unique material.

In the Lerf-Klinowski model of graphene epoxide [20], oxygen atoms are covalently bonded to two neighboring carbon atoms in graphene, forming an epoxy ring group. Optical image studies and energetic analysis show these epoxy groups like to form fault lines cooperatively, to lower their formation energy [18]. The adsorption of epoxy group was found to be able to break the $sp^2$ carbon bonds underneath and the process was discussed as oxygen-driven "unzipping" of graphitic materials [18, 23]. One the other hand, recent scanning probe microscopy identified periodic pattern of the oxygen atoms bound to the graphene sheet [19], where the oxygen lines are also aligned along the zigzag edges. Instead of breaking carbon-carbon (C-C) bond in the unzipping mechanism, the C-C bond length in the periodic structures is found to be very close to the value of pristine graphene [19]. Inspired by these studies, we proposed the graphene epoxide model, as depicted in Figure 1 to probe their structure and properties. In this model, the epoxy groups line up in *y* direction, and are patterned periodically along the *x*-direction (armchair). We characterize the structure of the graphene epoxide using the unit cell as shown in Figure 1. The character *width number* of the unit cell *w* is defined in unit of the graphene lattice period in the *x*-direction. As an example, the unit cell shown in Figure 1(a) and (b) has width $w = 2$. Following this definition, the unit cell with width number *w* has the chemical composition of one epoxy group and 4*w* carbon atom, i.e. $C_{4w}O$ ($w = 1, 2, …$) and is used to investigate the bulk properties of graphene oxides with a corresponding epoxidation density $1/(4w)$. It is also noticed that the epoxy group can be adsorbed on either side of the graphene, so there are different configurations for $C_{4w}O$, where the neighboring epoxy groups (*A* and *B* in Figure 1) can be adsorbed on the same ($C_{4w}O_{asym}$) or opposite side ($C_{4w}O_{sym}$) of the graphene (see insets in Figure 2).

*Methods*:

The structures of the graphene epoxide are determined by first principle calculations. In our work, plane-wave basis sets based density functional theory (DFT) was employed with local (spin) density approximation (LDA). We used the PWSCF code [24] for the calculation. The pseudopotential parameters were taken from Perdew-Zunger sets [25]. For all the results presented in this paper, energy cut-offs of 30 Rydberg and 240 Rydberg are used for plane-wave basis sets and charge density grids respectively. The settings have been verified to achieve a total energy convergence less than $10^{-5}$ Ry/atom. For variable-cell relaxation, the criteria for stress and force on atoms were set to be 0.01 GPa and 0.001 Ry/Å. Eight Monkhorst-Pack *k*-points were used in each periodic direction for Brillouin zone integration, which was qualified for the energy convergence criteria of 3 meV.

*Results and Discussion*:

The structures of graphene epoxide are obtained after variable-cell geometric optimization. The result shows that graphene sheet buckles around the epoxy groups and the distortion depends on the epoxide density $1/(4w)$, as summarized in Table 1. For $w <= 1$, there exists metastable *clamped* structure (Figure 1(c)) where the epoxy ring is close to an equilateral triangle. In these structures, the adsorption of oxygen atom stretches the underneath C-C bonds by 5% but doesn't break the $sp^2$ carbon network (Figure 3(b)). However this clamped phase is metastable. As we increase the lattice constant, the graphene epoxide collapses into a more stable *unzipped* phase with $sp^2$ bond underneath the oxygen atom broken, i.e. $E_{clamped} - E_{unzipped} = 0.86$ eV. A significant energy barrier between these two phases ($E_b$

= 0.58 eV for $w = 1$, and 1.7 eV for $w = 0.5$) makes the clamped phase considerably stable. For $w > 1$, the metastable clamped phase disappears, leaving the unzipped structure the only possible configuration. The angle C-O-C between two carbon-oxygen bonds in unzipped epoxy ring is about 150 degree, and the final C-C bond is broken with an extended distance of 2.5 Å (Figure 1c). The structure of the epoxy ring doesn't change much as the width $w$ increases further, because the strain energy induced by the local distortion decays quickly as the oxidation density decreases. Thus sparse epoxy lines could result in folds and wrinkles observed in graphene oxide [15]. Table 1 summarizes the structural properties of various graphite epoxides.

These results explain both the discrepant observations of the graphene oxide structure in [19] and [18]: the periodic epoxy lattice observed in resolved STM images has lattice constants 2.73 Å ($x$) and 4.06 Å ($y$) which is comparable with 2.84 Å ($x$) and 4.20 Å ($y$) for graphene sheet [19], this structure corresponds to the clamped phase with $w = 1$; while the "unzipping" mechanism for the epoxy line network is consistent with the unzipped phase observed here. The broken C-C bond in the epoxy group could give rise to the observable lines in optical microscopy [18].

The binding strength between the epoxy group and graphene sheet is quantified through the formation energy $E_f = E_{graphene\ epoxide} - (E_{graphene} + E_{oxygen}/2)$. $E_{graphene\ epoxide}$ and $E_{graphene}$ are the energy of graphene epoxide and pristine graphene. The reference energy value for oxygen atom is evaluated in isolated $O_2$ molecule spin-unpolarized state as $E_{oxygen}/2$. The results depicted in Figure 2 show that the unzipped structures bind stronger with the graphene. In clamped phase, the formation energy decreases as the oxidation density $1/(4w)$ increases as shown in Figure 2 and Ref. [16, 26], while for unzipped structures, it increases, because at elevated epoxy density the distortion of graphene is built up. In addition, asymmetric arrangement of the epoxy group on both sides of the graphene lowers the formation energy further (for unzipped structures). These asymmetric structures (Figure 2) help to keep the planar structure of the graphene between epoxy groups and the final structure is relaxed in a zigzag way.

However for large *w*, the asymmetric structure is stabilized in a unzipped phase and deviates from planar sheet configuration. The thickness of the zigzag structure is linearly proportional to *w*, which contradicts with the experimental evidence of graphene oxide thickness (around 7 Å) [14, 19]. When the width number *w* increases after 2, the distortion energy of graphene sheets gets able to be released more effectively, and the formation energy converges for large *w*'s. For *w* > 12, the formation energy approaches a value of $E_f$ = -1.06 eV. This significant binding strength makes the epoxy group hard to be removed in the reduction process and thus prohibit high-quality synthesis for pristine graphene materials from graphene exfoliate [15].

The formation of epoxy ring and breaking of covalent $sp^2$ bond imply remarkable effects on the mechanical property of graphene. To understand this effect, mechanical loading has been applied to the unit cell of both pristine graphene and graphene epoxide $C_8O$ (*w* = 2). The stress-strain relations are plotted in Figure 3. Stresses are defined by considering the sheet has a thickness of 3.34 Å, which is the inter-layer distance in graphite crystals. It is found that epoxy groups soften the structure significantly, especially in the compression deformation, because of the flexibility of pre-stretched C-C bond and foldable C-O-C angle. The epoxy ring is found to reduce the Young's modulus of graphene by 42.4% (from 1060 to 610 GPa) along *x*-direction and 9.63% (from 1061 to 960 GPa) in the *y*-direction. At the small strain the loading is born by the bending of C-O-C angle rather than stretching and compression of $sp^2$ bonds in graphene. In contrast to the huge reduction of Young's modulus, the tensile strength along *x*-direction almost doesn't change (111 GPa), and elastic limit are decrease from 0.20 to 0.18. During the tensile loading process along *x*-direction, the epoxy ring bends into the graphene sheet plane at first. Then the structure starts to break after the in-plane bonds are elongated to their limit (Figure 3). The charge density distribution (Figure 3(d)) shows the C-C bond starts to break at strain of 0.18. It is noticed that at the breaking point of pristine graphene and graphene epoxide, the C-C bond along the tensile direction are elongated by 21% and 23%. The close values suggest the strength of the sheets are mainly contributed by the $sp^2$ carbon bonds and explained why the tensile strength doesn't change much as the Young's modulus.

The binding of epoxy groups also perturbs the electronic structure of graphene epoxide by destroying the local π-electron states. Temperature dependent electrical measurements and Raman spectroscopy revealed semiconducting behavior of graphene oxides [8, 27]. In the clamped phase, because the C-C bond doesn't break, so the oxygen adsorbed dopes the graphene lattice, the hybridization between π-bands in graphene and oxygen 2p orbital opens a gap and introduce an localized oxygen $2p_z$ band as shown in Figure 4(b) [16, 26].

For unzipped graphene epoxide $C_2O$ ($w$ = 0.5), the oxygen atom lies in the graphene lattice and bridges two carbon atoms. The hybridization between π electrons in oxygen and carbon atoms results in a zero-gap band structure (Figure 4(c)) with bands crossing near the Fermi level. While for the unzipped structures with $w >= 1$, epoxy lines cut the graphene into zigzag-edged graphene nano-ribbons and block the π-electron state of graphene (Figure 4(d)). The semiconducting gaps opened are inversely proportional to width of separated graphene ribbons (Figure 4(a)). The relation reflects the particle-in-box nature of the mobile π-state between epoxy barriers and can be explained using the two-band model for graphene. The dispersion relation of π state in graphene is $E(\mathbf{k}) = \pm \hbar v_F |\mathbf{k} - \mathbf{k}_0|$, where $\mathbf{k}$ is the wave vector and $\mathbf{k}_0$ is the wave vector where the two bands cross. $v_F$ is the Fermi velocity. The energy gap is thus predicted to be $E_g \sim h v_F / w$ and is consistent with our calculation results. In the asymmetrically unzipped structures, the planar graphene structure between epoxy groups preserve the crossing of π and $π^*$ bands at Fermi level as shown in Figure 4(e).

It was recently discovered that hydrogen terminated graphene nano-ribbons, edge localized states provide half-metallicities and unique magnetic properties [5]. Electron orbital analysis shows that in the unzipped structures, the valence $v_1$ and conduction band $c_1$ near the Fermi level for $w >= 1$ are localized states at the epoxy group which resemble edge states in graphene nanoribbons (Figure 4(a)) [5, 28].

*Conclusion*:

In conclusion, we investigated graphene epoxide as an example of functional group engineered graphene materials, focusing on their mechanical and electronic properties at various oxidation conditions. For regularly pattern epoxy structures, two phases are revealed to have considerable binding strength in consistence with previous experimental observations: clamped structure where the oxygen adsorbed on $sp^2$ bond and unzipped structure where the epoxy binding breaks $sp^2$ bond. The first phase occurs at high oxidation density and forms regular lattice, while the second phase is more stable and results in line defects in graphene.

Mechanical properties and electronic structures are studied subsequently after the configuration has been determined. The bendable epoxy structure give rise to a half loss of Young's modulus, while leaves the tensile strength unaffected. The epoxy group also changes their band structures and give rise to various electronic properties, depending on the oxidation. These results suggest possible applications such as high strength composites and tunable electronic materials.

Our work laid the ground for the design and application of functional graphene-based materials through oxidation control methods. The understanding of the structural-property relation thus obtained is also critical for preparing graphene-related materials such as graphene oxide papers [29] through chemical exfoliation methods. Furthermore, with the ability to control the oxidation and reduction process, one can expect to fabricate tunable graphene-related nano-devices, such as chemical sensors [30] or patternable nano-electric circuits [6, 7].

**Reference:**

1.  Novoselov, K. S.; Geim, A. K.; Morozov, S. V.; Jiang, D.; Zhang, Y.; Dubonos, S. V.; Grigorieva, I. V.; Firsov, A. A., Electric Field Effect in Atomically Thin Carbon Films. *Science* **2004,** 306, (5696), 666-669.

2.  Novoselov, K. S.; Geim, A. K.; Morozov, S. V.; Jiang, D.; Katsnelson, M. I.; Grigorieva, I. V.; Dubonos, S. V.; Firsov, A. A., Two-dimensional gas of massless Dirac fermions in graphene. *Nature* **2005,** 438, (7065), 197-200.

3.  Novoselov, K. S.; Jiang, Z.; Zhang, Y.; Morozov, S. V.; Stormer, H. L.; Zeitler, U.; Maan, J. C.; Boebinger, G. S.; Kim, P.; Geim, A. K., Room-Temperature Quantum Hall Effect in Graphene. *Science* **2007**, 1137201.

4.  Lee, C.; Wei, X.; Kysar, J. W.; Hone, J., Measurement of the Elastic Properties and Intrinsic Strength of Monolayer Graphene. *Science* **2008,** 321, (5887), 385-388.

5.  Son, Y. W.; Cohen, M. L.; Louie, S. G., Half-metallic graphene nanoribbons. *Nature* **2006,** 444, (7117), 347-349.

6.  Xu, Z.; Zheng, Q.-S.; Chen, G., Elementary building blocks of graphene-nanoribbon-based electronic devices. *Applied Physics Letters* **2007,** 90, (22), 223115-3.

7.  Ci, L.; Xu, Z.; Wang, L.; Gao, W.; Ding, F.; Kelly, K.; Yakobson, B.; Ajayan, P., Controlled nanocutting of graphene. *Nano Research* **2008,** 1, (2), 116-122.

8.  Gómez-Navarro, C.; Weitz, R. T.; Bittner, A. M.; Scolari, M.; Mews, A.; Burghard, M.; K., K., Electronic Transport Properties of Individual Chemically Reduced Graphene Oxide Sheets *Nano Lett.* **2007,** 7, (11), 4.

9.  Eda, G.; Fanchini, G.; Chhowalla, M., Large-area ultrathin films of reduced graphene oxide as a transparent and flexible electronic material. *Nat Nano* **2008,** 3, (5), 270-274.

10. Tung, V. C.; Allen, M. J.; Yang, Y.; Kaner, R. B., High-throughput solution processing of large-scale graphene. *Nat Nano* **2008,** advanced online publication.

11. Eda, G.; Fanchini, G.; Chhowalla, M., graphene oxide as a transparent and graphene oxide as a transparent and flexible electronic material. *Nature Nanotechnology* **2008,** 3.

12. Berger, C.; Song, Z.; Li, T.; Li, X.; Ogbazghi, A. Y.; Feng, R.; Dai, Z.; Marchenkov, A. N.; Conrad, E. H.; First, P. N.; deHeer, W. A., Ultrathin Epitaxial Graphite: 2D Electron Gas Properties and a Route toward Graphene-based Nanoelectronics. *J. Phys. Chem. B* **2004,** 108, (52), 19912-19916.

13. Yang, X.; Dou, X.; Rouhanipour, A.; Zhi, L.; Räder, H. J.; Müllen, K., Two-Dimensional Graphene Nanoribbons *Journal of American Chemical Society* **2008,** 130, (13), 2.

14. Jeong, H.; Lee, Y. P.; Lahaye, R. J. W. E.; Park, M. H.; An, K. H.; Kim, I. J.; Yang, C.; Park, C. Y.; Ruoff, R. S.; H., L. Y., Evidence of graphitic AB stacking order of graphite oxides. *Journal of American Chemical Society* **2008,** 130, 5.

15. Stankovich, S.; Dikin, D. A.; Piner, R. D.; Kohlhaas, K. A.; Kleinhammes, A.; Jia, Y.; Wu, Y.; Nguyen, S. T.; Ruoff, R. S., Synthesis of graphene-based nanosheets via chemical reduction of exfoliated graphite oxide. *Carbon* **2007,** 45, (7), 1558-1565.

16. Boukhvalov, D. W.; Katsnelson, M. I., Modeling of Graphite Oxide. *Journal of the American Chemical Society* **2008,** 130, (32), 10697-10701.

17. McAllister, M. J.; Li, J.-L.; Adamson, D. H.; Schniepp, H. C.; Abdala, A. A.; Liu, J.; Herrera-Alonso, M.; Milius, D. L.; Car, R.; Prud'homme, R. K.; Aksay, I. A., Single Sheet Functionalized Graphene by Oxidation and Thermal Expansion of Graphite. *Chemistry of Materials* **2007,** 19, (18), 4396-4404.

18. Li, J.-L.; Kudin, K. N.; McAllister, M. J.; Prud'homme, R. K.; Aksay, I. A.; Car, R., Oxygen-Driven Unzipping of Graphitic Materials. *Physical Review Letters* **2006,** 96, (17), 176101-4.

19. Pandey, D.; Reifenberger, R.; Piner, R., Scanning probe microscopy study of exfoliated oxidized graphene sheets. *Surface Science* **2008,** 602, (1607).

20. Lerf, A.; He, H.; Forster, M.; Klinowski, J., Structure of Graphite Oxide Revisited. *The Journal of Physical Chemistry B* **1998,** 102, (23), 4477-4482.

21. Schniepp, H. C.; Li, J.-L.; McAllister, M. J.; Sai, H.; Herrera-Alonso, M.; Adamson, D. H.; Prud'homme, R. K.; Car, R.; Saville, D. A.; Aksay, I. A., Functionalized Single Graphene Sheets Derived from Splitting Graphite Oxide. *The Journal of Physical Chemistry B* **2006,** 110, (17), 8535-8539.

22. Cai, W.; Piner, R. D.; Stadermann, F. J.; Park, S.; Shaibat, M. A.; Ishii, Y.; Yang, D.; Velamakanni, A.; An, S. J.; Stoller, M.; An, J.; Chen, D.; Ruoff, R. S., Synthesis and Solid-State NMR Structural Characterization of 13C-Labeled Graphite Oxide. *Science* **2008,** 321, (5897), 1815-1817.

23. Alexandre, S. S.; Mazzoni, M. S. C.; Chacham, H., Edge States and Magnetism in Carbon Nanotubes with Line Defects. *Physical Review Letters* **2008,** 100, (14), 146801-4.

24. PWSCF (distributed in Quantum-Espresso package) is a community project for high-quality quantum-simulation software, based on density-functional theory, and coordinated by Paolo Giannozzi. See http://www.quantum-espresso.org and http://www.pwscf.org.

25. Rappe, A. M.; Rabe, K. M.; Kaxiras, E.; Joannopoulos, J. D., Optimized pseudopotentials. *Physical Review B* **1990,** 41, (2), 1227.

26. Ito, J.; Nakamura, J.; Natori, A., Semiconducting nature of the oxygen-adsorbed graphene sheet. *J. Appl. Phys.* **2008,** 103.

27. Gilje, S.; Han, S.; Wang, M.; Wang, K. L.; Kaner, R. B., A Chemical Route to Graphene for Device Applications. *Nano Letters* **2007,** 7, (11), 3394-3398.

28. Son, Y. W.; Cohen, M. L.; Louie, S. G., Energy gaps in graphene nanoribbons. *Physical Review Letters* **2006,** 97, (21), 216803-4.

29. Dikin, D. A.; Stankovich, S.; Zimney, E. J.; Piner, R. D.; Dommett, G. H. B.; Evmenenko, G.; Nguyen, S. T.; Ruoff, R. S., Preparation and characterization of graphene oxide paper. *Nature* **2007,** 448, (7152), 457-460.

30. Schedin, F.; Geim, A. K.; Morozov, S. V.; Hill, E. W.; Blake, P.; Katsnelson, M. I.; Novoselov, K. S., Detection of individual gas molecules adsorbed on graphene. *Nat Mater* **2007,** 6, (9), 652-655.

**Figures, Tables and Captions**

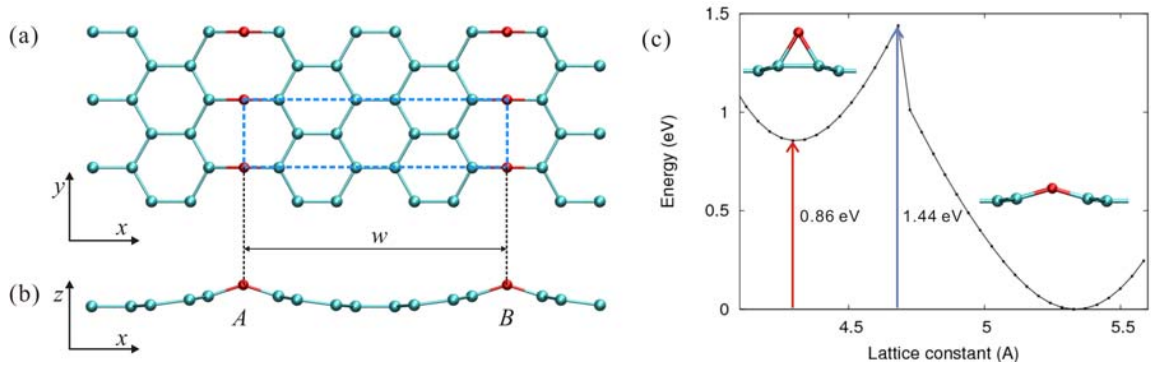

**Figure 1**: (a) and (b) Optimized structure of graphene epoxide. The unit cell (dashed rectangle) contains one oxygen atom (in red). The width number *w* of the unit cell is defined as multiples of the period of graphene unit cells. For the structure shown, *w* = 2. Epoxy group at neighboring sites A and B can be either in the same (*symmetric*) or opposite (*symmetric*) side of the graphene. (c) Energy landscape for graphene epoxide $C_4O$ (*w* = 1) as changing the lattice constant along *x*-direction, Insets shows its metastable clamped (left) and unzipped (right) structures, separated by a barrier of 0.58 eV.

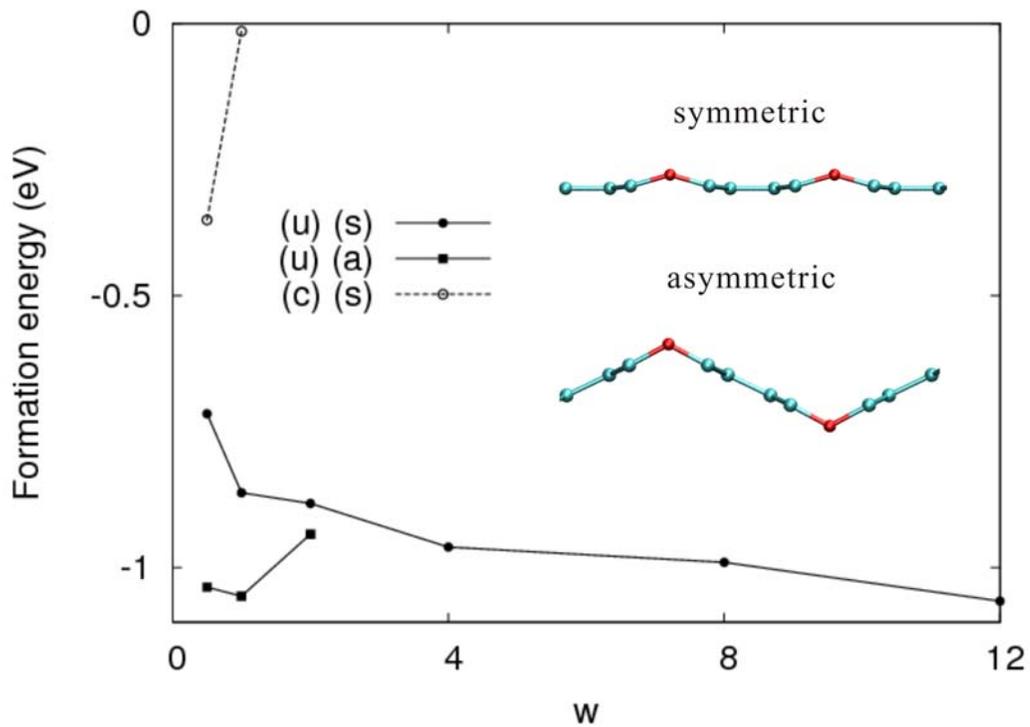

**Figure 2**: Formation energy of graphene epoxide with dependence on their width number $w$. The filled circles and squares stand for symmetrically/asymmetric unzipped structures. Open circles are symmetrically clamped structures, which disappears for $w > 1$. The insets show the optimized structure of symmetric and asymmetric structure for $C_4O$ ($w = 1$). (s), (a), (c), (u) in the figure denote symmetric, asymmetric, clamped and unzipped.

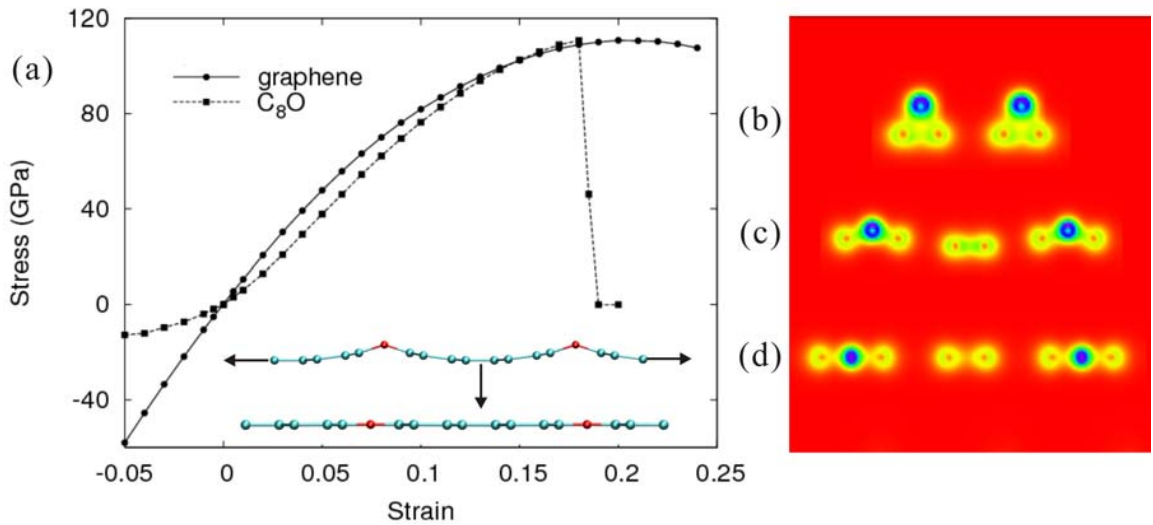

**Figure 3**: (a) Stress-strain relation of graphene and graphene epoxide. Insets: configuration change with strain from 0.0 to 0.18. (b)-(d) Charge density of the epoxy ring, plotted in the *x-z* plane along the *y*-line containing oxygen atoms. (b) shows the metastable clamped structure with C-C bond length at 1.5 Å (*w* = 1); (c) and (d) show the bending and flatten of C-O-C angle with strain 0.0 and 0.18 (at the breaking point) respectively for *w* = 2. Blue (red) color stands for high (low) electron density.

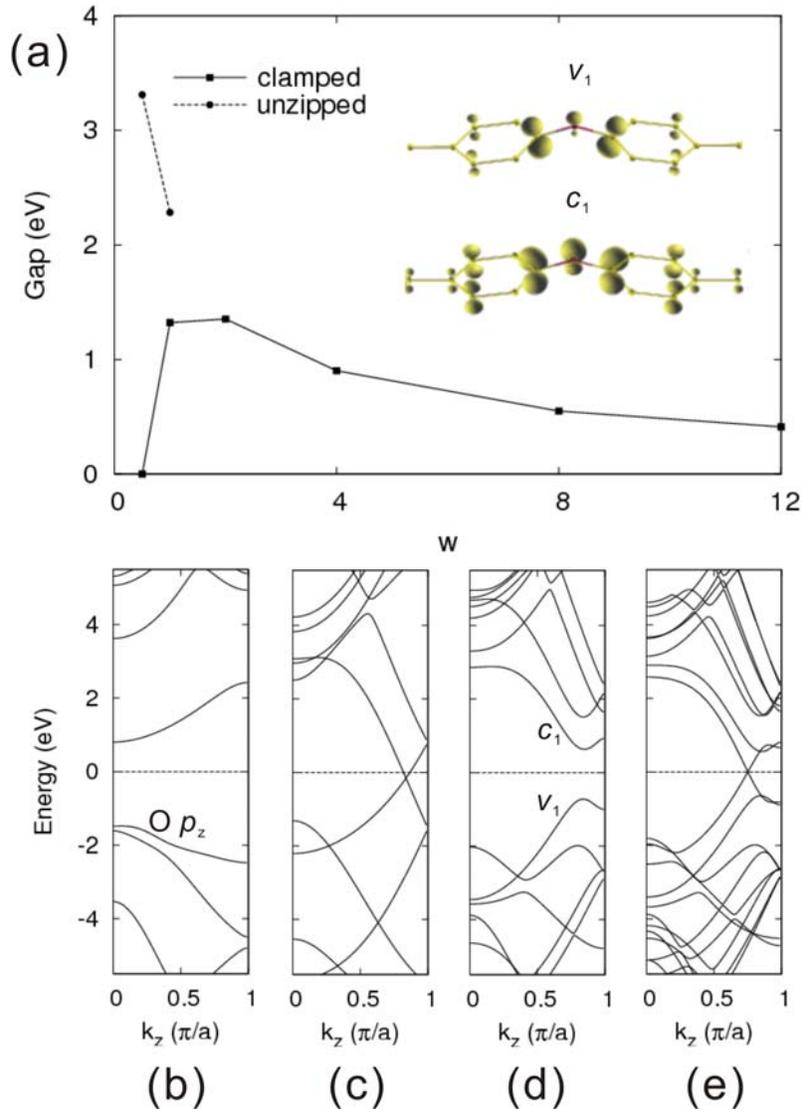

**Figure 4**: (a) Energy gaps of symmetrically and asymmetrically unzipped graphene epoxide with dependence on width $w$. Insets: charge distribution for the first valence ($v_1$) and conduction bands ($c_1$) of symmetrically unzipped structure. (b)-(d) show band structures for symmetrically clamped $C_4O_{sym}$ ($w = 1$), symmetrically unzipped $C_2O_{sym}$ ($w = 0.5$), $C_8O_{sym}$ ($w = 2$) and asymmetrically unzipped $C_8O_{asym}$ ($w = 2$) structures. The energy is shifted so that Fermi level $E_F = 0$ eV.

**Table 1: Structural properties of graphene epoxide with various oxygen density**

| $w$ | 0.5 (c)[1] | 0.5 (u)[1] | 1 (c) | 1 (u) | 2 (u) | 4 (u) | 8 (u) | 12 (u) |
|---|---|---|---|---|---|---|---|---|
| $d_{C-C}$ (Å)[2] | 1.50 | 2.58 | 1.49 | 2.55 | 2.53 | 2.51 | 2.51 | 2.53 |
| $A_{C-O-C}$ (°) | 63.9 | 153.6 | 62.3 | 148.2 | 144.6 | 142.1 | 142.9 | 146.2 |
| $\varepsilon_x$[3] | 0.083 | 0.533 | 0.018 | 0.262 | 0.123 | 0.050 | 0.027 | 0.019 |

[1]: (c) stands for clamped phase, (u) for unzipped phase

[2]: For graphene, $d_{C-C}$ = 1.408 Å. $d_{C-C}$ is the C-C distance underneath oxygen atom. $A_{C-O-C}$ is the C-O-C angle in epoxy group.

[3]: The *x*-direction strain is defined using lattice constants of pristine graphene as reference values.